\documentclass[
showpacs,
 amsmath,amssymb,
 aps,superscriptaddress,
 10pt,
twocolumn
]{revtex4-1}

\usepackage[usenames,dvipsnames]{color}
\usepackage{graphicx,textcase} 
\usepackage{dcolumn,overpic} 
\usepackage{bm,color}
\usepackage[T1]{fontenc}
\usepackage{ae,aecompl}
\usepackage{chngcntr,gensymb}
\usepackage[para]{threeparttable}
\usepackage{etoolbox}
\usepackage{lipsum}
\usepackage[bottom]{footmisc}
\usepackage{setspace}

\AtEndEnvironment{table*}{\vskip10pt}{}{}

\usepackage{url}
\usepackage{hyperref}
\hypersetup{colorlinks=true,breaklinks,linkcolor=blue,urlcolor=blue,citecolor=blue}

\begin{document}

\title{Magnetic 2D electron liquid at the surface of Heusler semiconductors}
\author{S.~Keshavarz}
\affiliation{Uppsala University, Department of Physics and Astronomy, Division of Materials Theory, Box 516, SE-751 20 Uppsala, Sweden}
\author{I.~Di~Marco}
\affiliation{Uppsala University, Department of Physics and Astronomy, Division of Materials Theory, Box 516, SE-751 20 Uppsala, Sweden}
\affiliation{Asia Pacific Center for Theoretical Physics, Pohang, Gyeongbuk 790-784, Korea}
\affiliation{Department of Physics, POSTECH, Pohang, Gyeongbuk 790-784, South Korea}
\author{D.~Thonig}
\affiliation{Uppsala University, Department of Physics and Astronomy, Division of Materials Theory, Box 516, SE-751 20 Uppsala, Sweden}
\author{L. Chioncel}
\affiliation{Theoretical Physics III, Center for Electronic Correlations and Magnetism, Institute of Physics, University of Augsburg, D-86135 Augsburg, Germany}
\affiliation{Augsburg Center for Innovative Technologies, University of Augsburg, D-86135 Augsburg, Germany}
\author{O.~Eriksson}
\affiliation{Uppsala University, Department of Physics and Astronomy, Division of Materials Theory, Box 516, SE-751 20 Uppsala, Sweden}
\affiliation{School of Science and Technology, \"Orebro University, SE-701 82 \"Orebro, Sweden}
\author{Y.~O.~Kvashnin}
\affiliation{Uppsala University, Department of Physics and Astronomy, Division of Materials Theory, Box 516, SE-751 20 Uppsala, Sweden}

\date{\today}

\begin{abstract}
Conducting and magnetic properties of a material often change in some confined geometries.
However, a situation where a non-magnetic semiconductor becomes both metallic and magnetic at the surface is quite rare, and to the best of our knowledge has never been observed in experiment. 
In this work, we employ first-principles electronic structure theory to predict that such a peculiar magnetic state emerges in a family of quaternary Heusler compounds. 
We investigate magnetic and electronic properties of CoCrTiP, FeMnTiP and CoMnVAl. For the latter material, we also analyse 
the magnetic exchange interactions and use them for parametrizing an effective spin Hamiltonian. According to our results,  magnetism in this material should persist at temperatures at least as high as 155 K.
\end{abstract}

\maketitle

\textit{Introduction.}
In several classes of materials, surfaces are known to exhibit properties distinctly different from those of the bulk. At quantum mechanical level, surface electronic states were first characterized almost a century ago, in the pioneering works of Tamm~\cite{tamm} and Shockley~\cite{PhysRev.56.317}. A direct probe of surface states became possible only much later, with the improvements in experimental techniques, as e.g. in angular resolved photo-electron spectroscopy (ARPES)~\cite{Ni-111,Cu-111,W-Mo-100,bcc-str-001}. The arising of surface physics led to the observation of several key phenomena. Materials characterized by a band gap in their bulk electronic structure can possess a metallic surface hosting a two-dimensional electron gas (2DEG), as e.g in SrTiO$_3$~\cite{Santander-Syro}, or the much celebrated topological edge modes~\cite{kane-mele}. Differences between surface and bulk electronic structures are also the foundation of interface-induced effects, as those observed when combining LaAlO$_3$  and SrTiO$_3$. The interface of these two insulators does not only host a peculiar 2DEG~\cite{Ohtomo-2002,Ohtomo-2004,Ahn}, but may also harbor a quasi-2D magnetic order driven by oxygen defects~\cite{pa.ko.12a,pa.ko.12b}. These studies associate the interface magnetism with off-stoichiometry modelled by artificially ordered 
supercells, sometimes limited to high defect concentrations~\cite{pe.pi.06,ja.ve.08}. 

Confined geometries have always been of primary importance for Heusler and half-Heusler compounds~\cite{graf2011simple}. For instance, understanding interfaces is crucial for building Heusler-based devices for giant and tunneling magnetoresistances~\cite{graf2011simple,ga.ma.06,ch.gr.11}. Interfaces have not only a practical importance, but may also host non-collinear spin ordering~\cite{fetzer2015probing,heter-igor} resulting into complex magnetic structures. 
Recently, materials like LaPtBi, LuPtSb, ScPtBi, YPdBi, and ThPtPb were shown to represent a new class of three-dimensional topological insulators~\cite{Xiao-2010}. Furthermore, a new Weyl system was demonstrated to exist in a family of magnetic Heuslers~\cite{Wang-2016}.

Due to the high tunability and versatility of Heusler compounds, new phenomena are expected to emerge when more and more materials are suggested and synthesized. Precisely, in this letter we predict that a relatively unexplored (but already existing) family of quaternary Heusler compounds hosts a rare, if not unique, phenomenon, where a semiconducting and non-magnetic bulk coexists with a metallic surface with robust magnetism. By means of first-principles electronic structure calculations, we show this peculiar state to emerge in CoMnVAl, CoCrTiP and FeMnTiP. Using calculated exchange interactions, we illustrate that the magnetic long-range order at the surface is stable up to sizeable temperatures, which is a fundamental prerequisite for experimental verification. We discuss the results both in terms of fundamental understanding of this new class of electronic structures, as well as in terms of technological aspects, that may emerge when spin degrees of freedom enter semiconductor-based electronic devices~\cite{Min-2006, Jansen-2007, Appelbaum-07, Jonker-07, Hamaya-2009}.

To the best of our knowledge, the coexistence of a semiconducting and non-magnetic bulk with a metallic and magnetic surface has never been observed in experiment. 
Here we identify a promising prototype of such materials, which are also characterized by a spin-polarized 2D electron liquid emerging at the surface.
It is important to stress that differentiating liquid and gas is not only a naming convention. Localized $3d$ orbitals, which give rise to the surface metallicity and magnetism here, are likely to harbour strong electronic correlations, which may lead to a plethora of exotic effects. Half-metallic Heusler compounds are known to host a wide array of such features. For example, genuine many-body states, named non-quasiparticle states (NQP), may appear in half-metallic ferromagnets, leading to a reduction of spin polarization~\cite{ed.he.73,ir.ka.90,Katsnelson-review08,ch.sa.08}. Another example is given by the mass enhancement and non-Fermi liquid behavior observed in Fe$_2$VAl and related materials~\cite{graf2011simple}. These effects, as well as others, are expected to emerge in the reported systems, due to the considerable localization and reduced dimensionality. 

\textit{Theory and methods.}
Density functional theory (DFT) calculations were carried out using the full-potential linear muffin-tin orbital (FP-LMTO) code RSPt~\cite{rspt-web,rspt-book}. 
We used the generalized gradient approximation (GGA), in the formulation of Perdew, Burke, and Enzerhof~\cite{pbe}.
The crystal structure of the considered class of Heusler semiconductors is described by a chemical formula XX$'$YZ, where X, X$'$ and Y are transition metals and Z is an $sp$ element.
The surface was modelled in a supercell geometry consisting of four unit-cells stacked along the [001] direction. This means that eight (or nine) XX$'$ layers mediated by eight alternating YZ layers were considered~\footnote{We have considered both symmetric and asymmetric slabs. The former one consists of 17 layers, which prevents the formation of an electric dipole moment. On the other hand, it does not have an integer number of electrons per atom, which may lead to a spurious metallicity. To make sure that the metallicity is intrinsic and not an artifact of a finite-size model, additional calculations were made for an asymmetric 16-layer slab of CoMnVAl. Nearly identical density of states were obtained, which is a strong evidence for the emergence of genuine metallic surface states.}.
A vacuum of 25~\AA~ thickness has been added to ensure no interaction between the surfaces. 
The computed equilibrium lattice parameters for all considered materials were taken from Ref.~\cite{main-blugel}.
Selected calculations of the FeMnTiP slabs were performed with a full relaxation of the atomic positions. The effects of the relaxation were found to be minor, and therefore we restricted our most extensive analysis, presented in the following, to the unrelaxed structures.

In addition to standard GGA, we also performed calculations where the effects of strong electron-electron repulsion were included explicitly, at the level of a mean-field DFT+$U$ approach~\cite{an.ar.97,gr.ma.12}. 
For simplicity, we adopted the same values of the Coulomb interaction parameters for the $d$-states of all X and X$'$ elements, namely {$U=2$ eV} and {$J=0.8$ eV}.
These values are situated in between the partly and fully screened estimates~\cite{u-value}, obtained for similar compounds using the constraint random phase approximation (cRPA)~\cite{cRPA}.

The thermal stability of the predicted magnetic long range order at the surface was investigated by means of an effective Heisenberg model. The latter is described by a Hamiltonian $\hat{H}=-\sum_{i\neq j} J_{ij}\vec{e}_i\cdot\vec{e}_j$, where $J_{ij}$ is the exchange interaction between the spins located at the site $i$ and $j$, and $\vec{e}$ is a unit vector along the spin direction at the corresponding site.
The $J_{ij}$'s were extracted from a GGA(+$U$) calculation by means of the magnetic force theorem~\cite{jij, jij-ldapp-2000, Jijs-in-rspt}.
Based on these values of the $J_{ij}$'s, we calculated the ordering temperature ($T_c$) by means of a classical Monte Carlo (MC) simulation, as implemented in the UppASD code~\cite{uppasd,uppasd-book}.
The adiabatic magnon spectra (AMS) were calculated for a considered magnetic ground state via Fourier transforms of the $J_{ij}$'s. 

\begin{figure*}[!t]
\includegraphics[width=2.0\columnwidth]{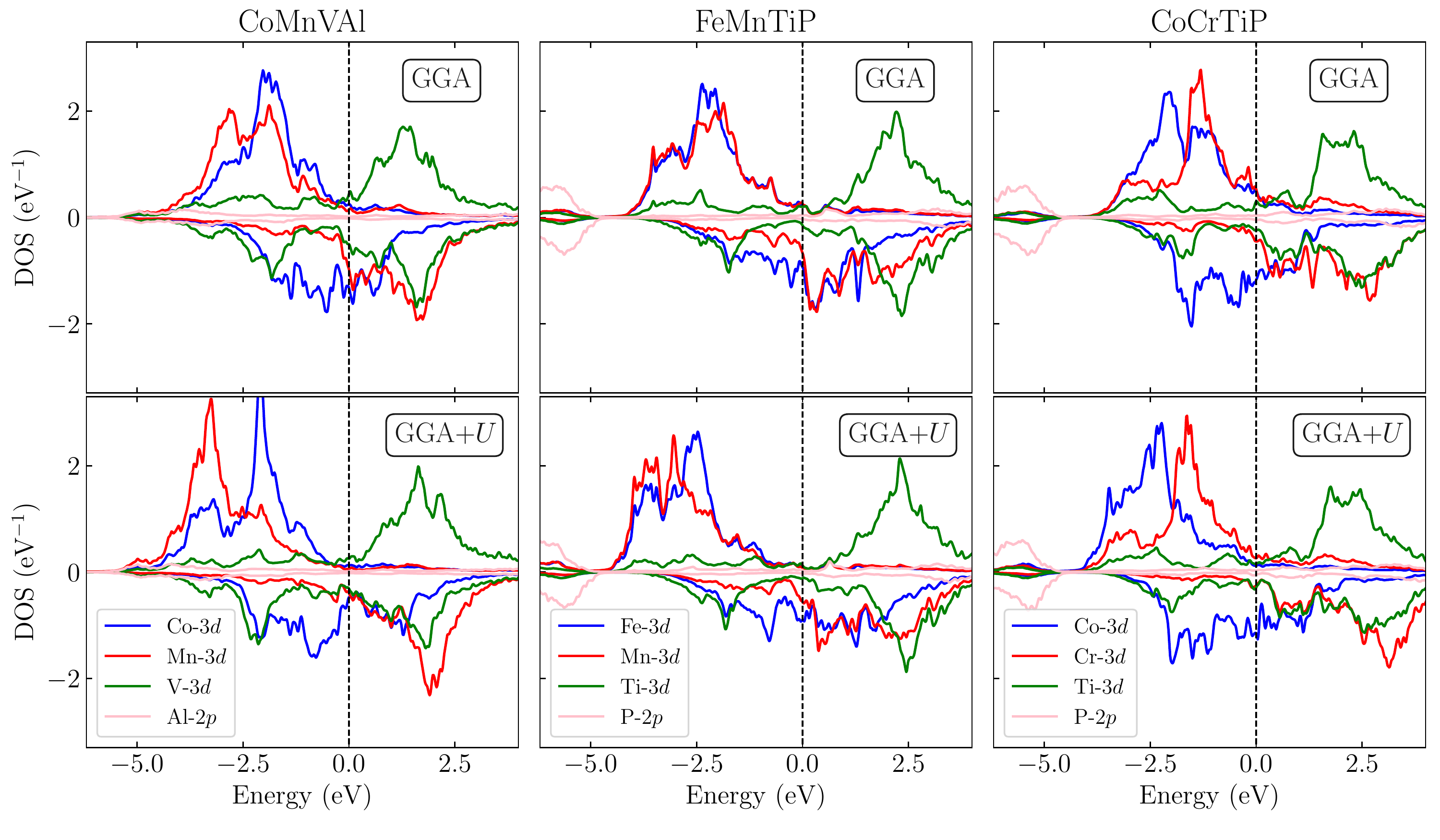} 
\caption{Projected density of states of the surface atoms (blue and red lines) and the subsurface atoms (green and pink lines) obtained with GGA (top panel) and GGA+$U$ (bottom panel). Fermi level is indicated by the dashed line.}
\label{fig:dos-unrel}
\end{figure*}

\textit{Results and Discussion.}
We first focus on the physical properties of three Heusler semiconductors: CoMnVAl, CoCrTiP and FeMnTiP.
According to the Slater-Pauling rule~\cite{slater,pauling}, Heusler compounds with 24 valence electrons are expected have zero net moment. They can either represent compensated ferrimagnetic half-metals or semiconductors with antiferromagnetic or non-magnetic orders~\cite{oz.sa.13,PhysRevLett.74.1171}.
The results of our calculations indicate that the selected materials belong to the latter category, in agreement with previous literature~\cite{main-blugel}. The calculated band gaps in GGA (GGA+$U$) are 0 (0.2) for CoMnVAl, 0.15 (0.3) for CoCrTiP and  0.5 (0.8) for FeMnTiP, where all values are given in eV.
The previous values show that GGA predicts CoMnVAl to be metallic, which is remedied by the inclusion of the $U$ term.
The situation is reminiscent of Co$_2$FeSi, where on-site correlations are also needed to obtain the true half-metallic solution~\cite{co2fesi-1,co2fesi-2}.
CoCrTiP and FeMnTiP are already semiconducting in standard GGA, but the band gaps are further enhanced when $U$ is added.

Next, we focus on the finite-size slabs, starting with the analysis of their spectral properties.
The projected density of states (DOS) for the atoms closest to the surface is reported in Fig.~\ref{fig:dos-unrel}, for all three considered systems. 
The DOS of the deeper-lying layers quickly converges to that of bulk materials, as shown in the Supplementary Material (SM) for CoMnVAl~\cite{SM}.
The innermost layers of the 16-layer slab are indistinguishable from the bulk, with perfect spin-degenerate bands, as was also reported in Ref.~\onlinecite{heter-igor}. 
The fundamental feature of Fig.~\ref{fig:dos-unrel} is that all systems are characterized by magnetic and metallic DOS at the surface and sub-surface layers.
Since the surfaces have a XX$'$-termination, the electronic states crossing the Fermi level primarily originate from the relatively localized $3d$ orbitals of the transition metals.
These localized surface states are the ones responsible for the arising of a finite magnetization.

The projected magnetic moments in the surface and sub-surface layers are reported in Table~\ref{tab:table1}.
Transition metal atoms at the surface exhibit pronounced magnetic moments, which are aligned ferromagnetically with each other. 
The Y atoms (V or Ti) belonging to the sub-surface layer show weak magnetic moments, which are anti-parallel to the large moments at the surface, thus resulting in a ferrimagnetic order.
The third layer below the surface is nearly non-magnetic, showing moments that are 10 to 300 times smaller than those at the surface. Going deeper into the slab, the bulk properties are recovered.

We note that the inclusion of the strong electron-electron repulsion via the local $U$ term does not change the qualitative picture obtained in plain GGA. In the projected DOS of Fig.~\ref{fig:dos-unrel}, the localization induced by the $U$ term leads to slightly narrower bands and larger spin splittings. Accordingly, the magnetic moments reported in  Table~\ref{tab:table1} are moderately enhanced in the DFT+$U$ approach, when compared to the values obtained in GGA. 

\begin{table}[!h]
\caption{\label{tab:table1} Calculated site-projected spin moments ($\mu_B$) for the considered Heusler compounds of XX$'$YZ family. The GGA+$U$-derived results are shown in parentheses. The indices refer to the layer numbers from the surface: surface (1), subsurface (2) and sub-subsurface (3). }
    \begin{tabular}{c|c|c|c}
            & CoMnVAl & CoCrTiP  & FeMnTiP  \\
            \hline
           X$_1$  &  1.36  (1.48)  &      1.11 (1.50)  &   2.31  (2.52)  \\  
           X$'_1$  &  3.37  (3.66)  &      3.02 (3.31)  &    3.14 (3.50)  \\ 
           Y$_2$   &  -0.51 (-1.03)    &    -0.12 (-0.18) &    -0.31 (-0.41) \\ 
           X$_3$  &  0.14  (0.16)   &     0.20 (0.14)   &    0.14  (0.08) \\ 
           X$'_3$  &  0.11  (0.26)    &    0.01  (-0.08)  &     0.02  (0.06)  \\ 
    \end{tabular}
\end{table}

In general, the surface of a non-magnetic material can become magnetic because of a reduced number of neighbouring atoms, which leads to a narrower bandwidth and the fulfillment of the Stoner criterion. In fact, calculations for the non-magnetic phase reveal that the surface DOS shows high peaks at the Fermi level, due to the aforementioned surface states. Thus, the Stoner criterion is fulfilled, and the spin-degeneracy of the bands is removed to lower the total energy (see SM~\cite{SM} for more details). 
In addition to the Stoner mechanism, this class of materials exhibit another peculiarity, which is the Slater-Pauling rule. The latter dictates the lack of magnetism in the bulk, as a consequence of having 24 valence electrons in the unit cell. However, this constraint does not hold any more at the surface, due to the dangling bonds caused by the reduced dimensionality. Therefore, it is almost natural for magnetism to arise. An analogous effect was previously reported for Heusler ferromagnets, whose surfaces were shown to exhibit a strongly modified electronic structure, lacking the half-metallicity of the bulk~\cite{Galanakis_2002}.

The previous analysis illustrates that magnetism and metallicity are basically confined to the surface and sub-surface layers. Both features are due to localized surface states of $3d$ character, which are expected to exhibit a substantial electron-electron interaction. Therefore, we refer to the observed situation as the formation of a (quasi) 2D electron liquid, to be distinguished from the 2DEG observed at certain interfaces, as discussed in the introduction. 

\begin{figure}
\includegraphics[width=\columnwidth]{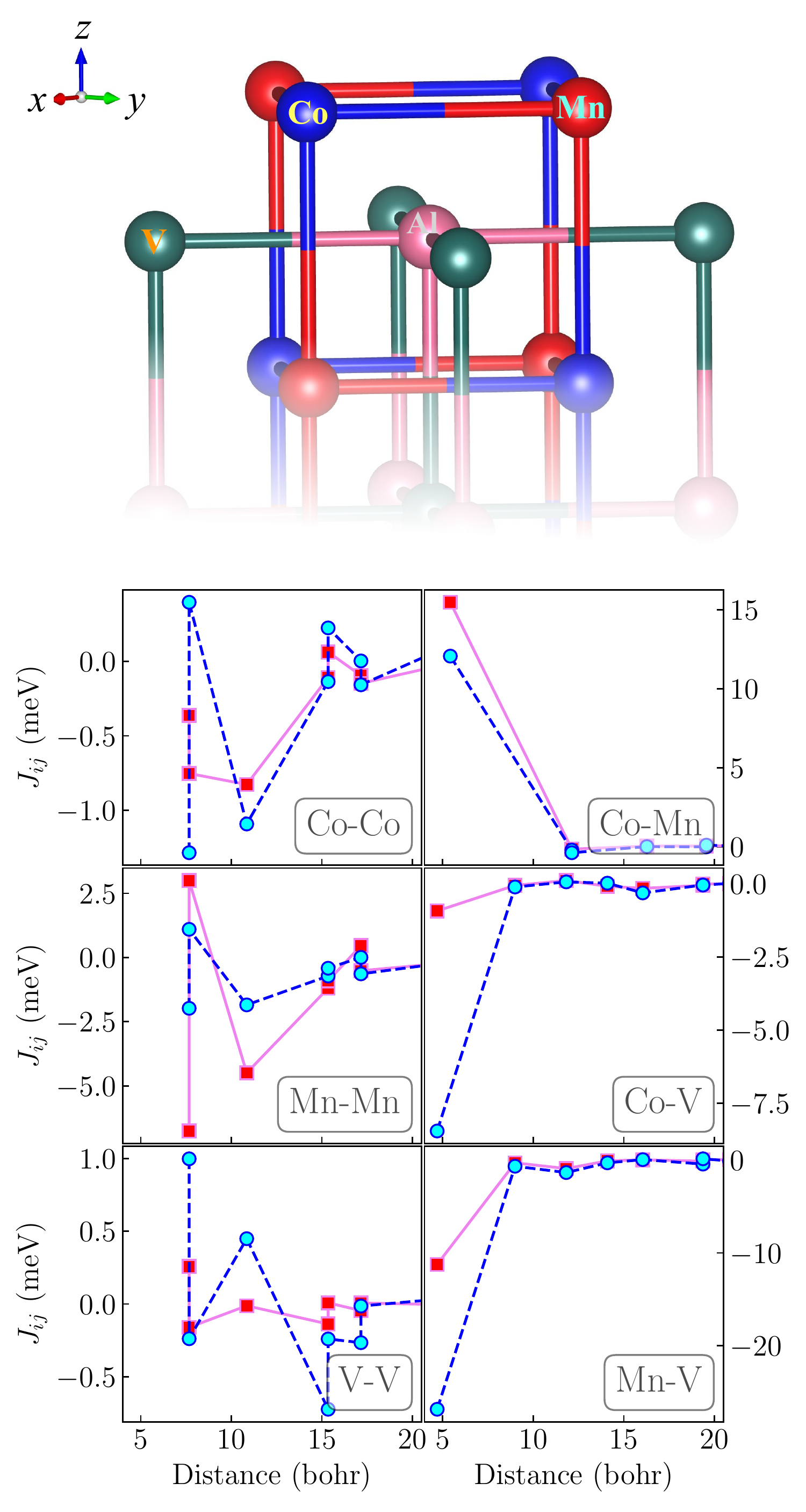} 
\caption{Top panel: Structure of the CoMn-terminated slab of CoMnVAl. Bottom panel: Inter-atomic exchange parameters $J_{ij}$'s as a function of the distance between the atoms. The results obtained with GGA (GGA+$U$) are shown with red squares (cyan circles).}
\label{fig:jijs}
\end{figure}

To get a deeper insight into the origin of the magnetic order at the surface, we have calculated the inter-atomic exchange parameters $J_{ij}$'s for all three systems. The calculated magnetic interactions are qualitatively similar in all cases, and therefore the following discussion will be focused only on CoMnVAl. The structure of the considered slab as well as the calculated $J_{ij}$'s are reported in Fig.~\ref{fig:jijs}. The important couplings are those involving Co, Mn and V atoms that are closest to the surface.
The interactions between other atoms were not computed, since the magnetic moments are too small and of induced nature, which makes the application of the magnetic force theorem doubtful.
The dominant interactions correspond to the nearest-neighbour (NN) Co-Mn, Co-V and Mn-V bonds.
Their signs are perfectly consistent with the ground state magnetic order, where surface and sub-surface spins are antiparallel to each other.
Taking the $U$ term into account does not change this picture, but results into an enhancement of all couplings, except the one between Co and Mn. 
The largest variation is observed for the interactions involving V atoms, and is partly induced by the large increase of the V moment (see Table~\ref{tab:table1}). 
The NN couplings between the atoms of the same element (e.g. Mn-Mn) are strongly bond-dependent.
This is a manifestation of the $C_{2v}$ symmetry of the slab, which makes the $x$ and $y$ direction nonequivalent.
In practice, this implies that the corresponding bonds have different environments and two of them have Al atoms positioned below and for the other two there are V atoms. 
These vastly different exchange paths affect and the sign and the magnitude of the $J_{ij}$.
The antiferromagnetic NN Co-Co, Mn-Mn and V-V interactions are, in fact, frustrated. 
They are reduced in GGA+$U$ calculation, which enhances the overall stability of the ferrimagnetic order, as will be shown below.

\begin{figure}[!t]
\includegraphics[width=\columnwidth]{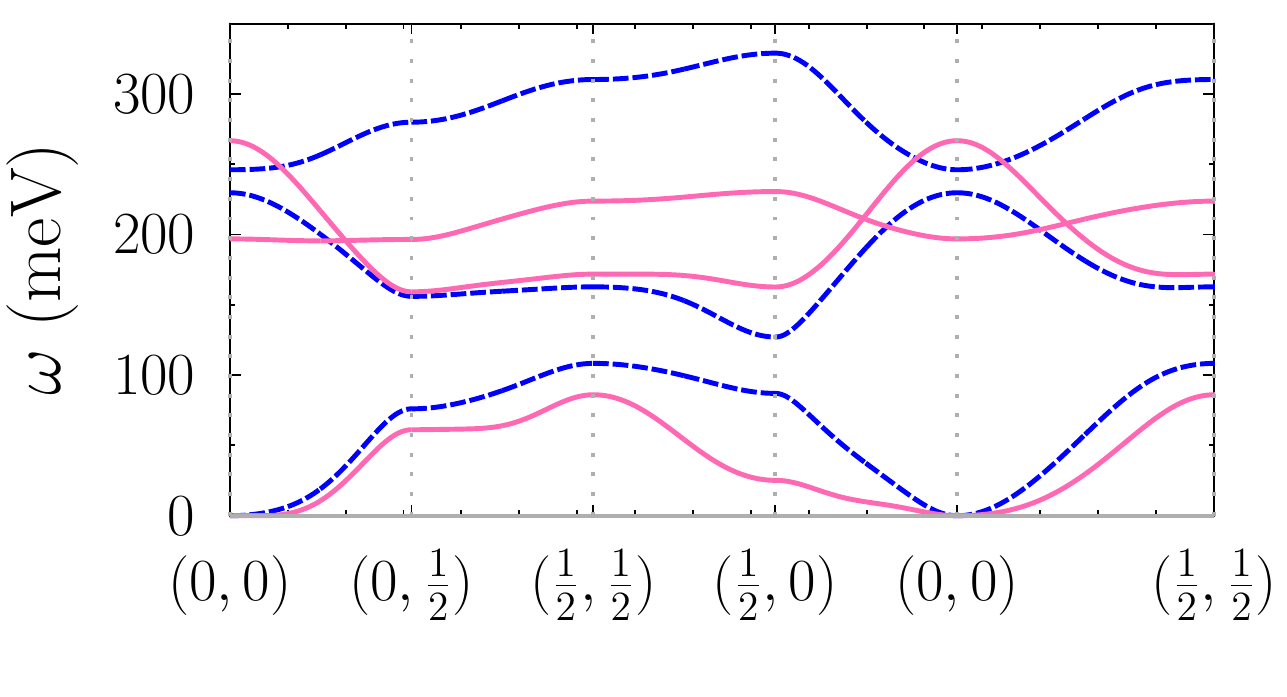} 
\caption{Adiabatic magnon spectrum of the CoMn-terminated CoMnVAl slab along the high-symmetry lines in the plane of the surface. Solid (dashed) lines correspond to GGA(+$U$)-derived data.}
\label{fig:tc-ams}
\end{figure}

Electronic structure calculations suggest the formation a spontaneous surface magnetization at zero temperature. To assess the stability of the magnetic state at finite temperature, we performed MC simulations for the parameterized Heisenberg model. 
Ordering temperatures were identified from the specific heat. For CoMnVAl, $T_c$ is predicted to be in the range between 155 and 290 K,
as obtained in GGA and GGA+$U$ respectively.
Further information can be obtained through the AMS. The spectrum of CoMnVAl, reported in Fig.~\ref{fig:tc-ams}, shows that the predicted reference state is indeed stable.
Again we see the manifestation of the $C_{2v}$ symmetry, since the dispersion at ($\frac{1}{2}$,0) and (0,$\frac{1}{2}$) points is clearly different.
The softest magnons are found for the case of GGA calculation along ($\frac{1}{2}$,0) direction, which is primarily caused by the frustrated NN Mn-Mn couplings.
The overall stability of the ferrimagnetic state is drastically increased in the GGA+$U$ calculations, as revealed by the larger spin-stiffness constant in the AMS.
This result and the aforementioned difference in $T_c$ are both consistent with enhanced $J_{ij}$'s and suppressed frustration, obtained in GGA+$U$ with respect to GGA.
Due to localized nature of the surface-derived states, correlation effects beyond DFT are expected to be important.
Thus, the actual magnetic and electronic properties should be closer to those obtained with GGA+$U$.

\textit{Conclusion and outlook.}
We have demonstrated that XX$'$YZ Heusler materials, which are semiconducting in the bulk, host a metallic and ferrimagnetic (quasi) 2D electron liquid at their [001] XX$'$-terminated surface.
The ordering temperatures are expected to be within experimental range and maybe even close to room temperature, as predicted for CoMnVAl. 
Our predictions can be verified through several surface-sensitive experimental techniques. 
The magnetically ordered state appearing at reduced temperatures can be investigated using x-ray circular dichroism~\cite{PhysRevLett.70.694, PhysRevLett.75.152} or spin-polarized scanning tunnelling microscopy/spectroscopy~\cite{RevModPhys.81.1495}.
Metallicity at the surface can unambiguously be verified using ARPES with varied photon energies\cite{STROCOV2018100}.
Finally, surface magnon states can potentially be measured with spin-polarized electron energy loss spectroscopy~\cite{PhysRevLett.102.177206} and then compared to our predicted spectrum.

The peculiar magnetic state predicted in this work has a potential of being important for technological applications.
Having an intrinsic metallic magnetism at the surface of a semiconducting interior implies that these materials may provide a practical realization of a magnetic tunnel junction or a spin-injector within a single entity.
Overall, these systems seem very promising for the growing field of spintronics.

Our discovery requires further theoretical and experimental investigation.
It is interesting to study the effect of antisite disorder, which might lead to the appearance of further exotic magnetic states.
The spin-orbit coupling has been neglected in the present analysis and its inclusion might lead to new interesting features of these materials, such as non-trivial topology, Dzyaloshinskii-Moriya interactions at the surface, and an enhanced magnetocrystalline anisotropy.
Finally, we speculate that the dynamical correlation effects, which are relevant for Heusler half-metals~\cite{Katsnelson-review08}, might be even more interesting for the surface-derived states discussed here. 

\begin{acknowledgments}
L.C. gratefully acknowledges the financial support provided by the Augsburg Center for Innovative Technologies, and by the Deutsche Forschungsgemeinschaft (DFG, German Research Foundation) - Projektnummer 107745057 - TRR 80/F6.
O.E. acknowledges support from the Swedish Research Agency, the Knut and Alice Wallenberg Foundation, the Foundation for Strategic Research, The Swedish Energy Agency and eSSENCE.
The computer simulations are performed on computational resources provided by NSC allocated by the Swedish National Infrastructure for Computing (SNIC). 
\end{acknowledgments}

\end{document}


\title{Supplemental material for Magnetic 2D electron liquid on the surface of Heusler semiconductors}
\author{S.~Keshavarz$^{1}$, I.~Di~Marco$^{1,2}$,  D.~Thonig$^{1}$, L. Chioncel$^{3,4}$, O.~Eriksson$^{1,5}$, and Y.~O.~Kvashnin$^{1}$}
\affiliation{$^1$Uppsala University, Department of Physics and Astronomy, Division of Materials Theory, Box 516, SE-751 20 Uppsala, Sweden}
\affiliation{$^2$Asia Pacific Center for Theoretical Physics, Pohang, Gyeongbuk 790-784, Korea}
\affiliation{$^3$Theoretical Physics III, Center for Electronic Correlations and Magnetism, Institute of Physics, University of Augsburg, D-86135 Augsburg, Germany}
\affiliation{$^4$Augsburg Center for Innovative Technologies, University of Augsburg, D-86135 Augsburg, Germany}
\affiliation{$^5$School of Science and Technology, \"Orebro University, SE-701 82 \"Orebro, Sweden}

\maketitle


Here we show the projected density of states (DOS) in CoMnVAl, obtained in various computational setups.
The results of non-magnetic GGA calculations (left panel) reveal high density of states at the Fermi level ($E_F$). 
It primarily originates from the localized Mn and Co $3d$ orbitals located in the surface layer. 
Sub-surface states already shows a substantially reduced DOS at this energy and the bulk electronic structure properties are quickly recovered, as one approaches the innermost layer.

Thus, it is clear that the surface states show the tendency to an instability of the non-magnetic solution. 
The latter results in the emergence of pronounced magnetic moments in the spin-polarized GGA calculations, as seen in the results shown in the middle panel of Fig.~S1.
Strong exchange splitting of the Mn and Co states is apparent in the magnetic calculation.
Even though the splitting causes the reduction of the density of spin-up states near the $E_F$, a substantial amount of spin-down states remains there.

Taking into account Coulomb $U$ term leads to a further enhancement of the spin polarization (see right panel of Fig.~S1) and consequently pushes the $3d$ states away from the $E_F$.
As a result, a combined effect of spin-polarization and strong local interactions, the DOS at the $E_F$ is reduced, but remains finite.
The metallic bands are confined to the first few layers closest to the surface of the material.

\begin{figure}[!h]
\includegraphics[angle=0,width=180mm]{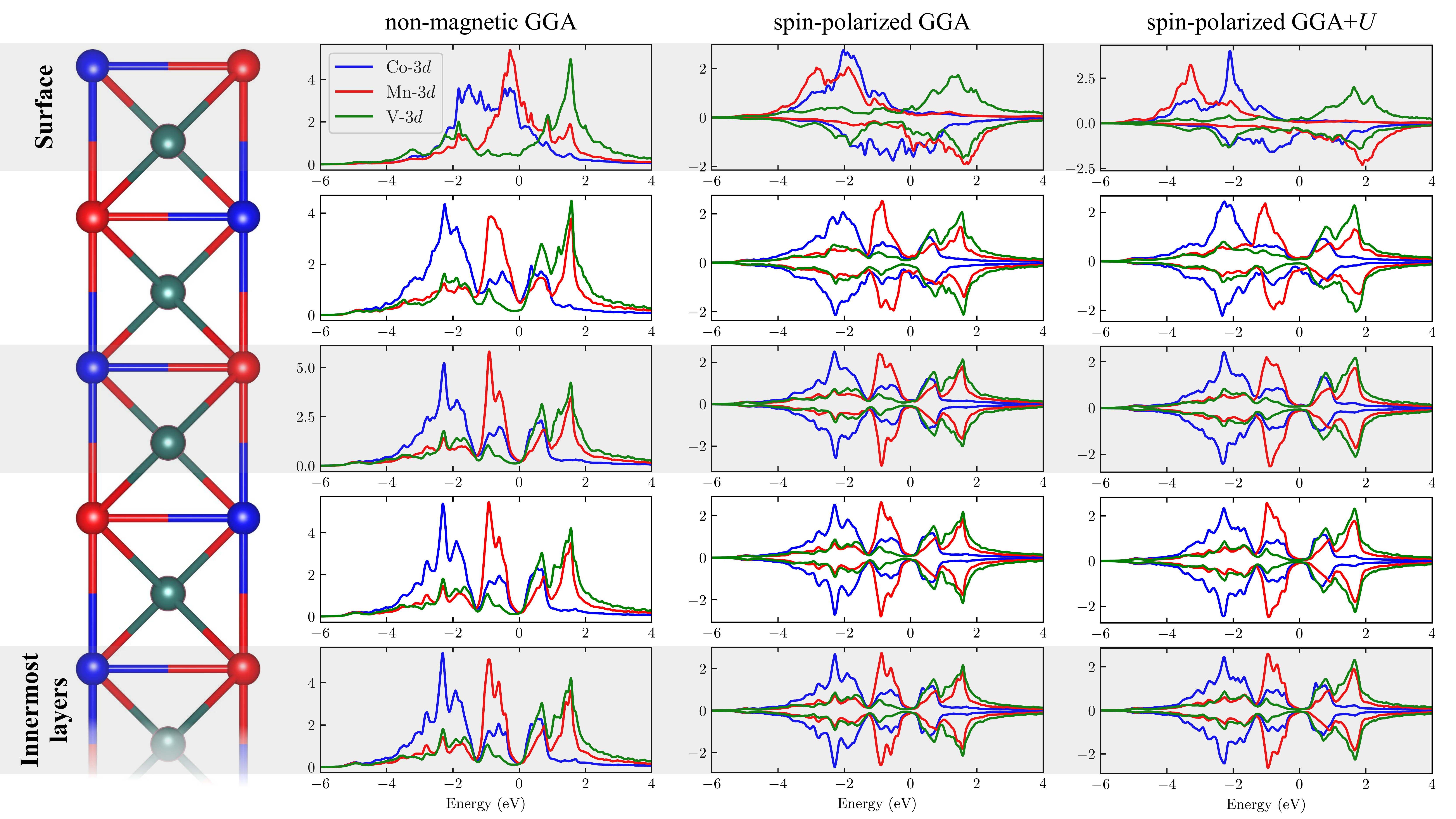}
\caption{Calculated layer-resolved density of states projected onto $3d$ states of transition metals in the slab of CoMnVAl.}
\label{PDOS}
\end{figure}